\documentclass[aps,pra,floatfix,twocolumn,superscriptaddress]{revtex4-1}
\usepackage{amsmath,amssymb}
\usepackage{graphicx}
\usepackage{epsfig}
\usepackage{psfrag}
\usepackage[usenames]{color}
\usepackage{xcolor}
\usepackage{hyperref}

\newcommand*{\colorboxed}{}
\def\colorboxed#1#{%
  \colorboxedAux{#1}%
}

\begin{document}

\title{Acoustoelectric effect in two-dimensional Dirac materials\\ exposed to Rayleigh surface acoustic waves}

\author{K.~Sonowal}
\affiliation{Center for Theoretical Physics of Complex Systems, Institute for Basic Science (IBS), Daejeon 34126, Korea}
\affiliation{Basic Science Program, Korea University of Science and Technology (UST), Daejeon 34113, Korea}

\author{A.~V.~Kalameitsev}
\affiliation{Rzhanov Institute of Semiconductor Physics, Siberian Branch of Russian Academy of Sciences, Novosibirsk 630090, Russia}

\author{V.~M.~Kovalev}
\affiliation{Rzhanov Institute of Semiconductor Physics, Siberian Branch of Russian Academy of Sciences, Novosibirsk 630090, Russia}
\affiliation{Novosibirsk State Technical University, Novosibirsk 630073, Russia}

\author{I.~G.~Savenko}
\affiliation{Center for Theoretical Physics of Complex Systems, Institute for Basic Science (IBS), Daejeon 34126, Korea}
\affiliation{Basic Science Program, Korea University of Science and Technology (UST), Daejeon 34113, Korea}

\begin{abstract}
We study the acoustoelectric effect in two-dimensional materials like transition metal dichalcogenide monolayers located on a non-piezoelectric substrate and exposed to the Rayleigh surface acoustic waves.
We investigate the behavior of the Hall component of the electric current density which appears due to the  trigonal warping of the valleys in k-space.
We calculate the spectrum of the current density and study its dependence on the electron effective lifetime and density in the sample. We distinguish between the drift and diffusive components of the current and show, which components turn out predominant. Furthermore, we compare the effect of the Rayleigh and Bluestein-Gulyaev acoustic waves, which appear if the sample is located on a piezoelectric substrate.
\end{abstract}

\date{\today}

\maketitle

\section{INTRODUCTION}
In the recent years, transition metal dichalcogenides (TMDCs)   have been attracting a great deal of attention in both the theoretical and experimental research~\cite{Radisavljevic2011, Sundaram2013}.
TMDCs represent a sub-class of two-dimensional (2D) Dirac materials that lack inversion symmetry and possess hexagonal lattice structure similar to that of graphene~\cite{Wang2012}. The presence of the valley degree of freedom and strong spin-orbit coupling makes them a promising platform for applications in the fields of valleytronics~\cite{Schaibley2016} and spintronics~\cite{Zibouche2014} .
Furthermore, the study related to the interaction of TMDCs with light fields have unravelled intriguing physical phenomena, which also makes these materials  potential candidates for optoelectronic devices~\cite{Mkrtchian2019, Have2019}.
However, the study of interaction of TMDCs with surface acoustic waves (SAWs) are still in their nascent stage~\cite{Kalameitsev2019,Sukhachov2020, RefOurAMEWeiss} and, to the best of our knowledge, the experiments are still lacking.
All this makes the physical implications of effects resulting due to the propagation of SAWs in TMDCs  an interesting and noteworthy field of research.

There exist several physical mechanisms of interaction between SAWs and the electrons of a two-dimensional electron gas (2DEG) in the sample.
In particular, in the case of the piezoelectric mechanism, when the 2DEG is located on a piezoelectric substrate, the interdigital transducers (IDTs) create the Bleustein-Gulyaev (BG) acoustic waves, which cause the drag of electrons.
This effect has been addressed in TMDCs~\cite{Kalameitsev2019}.

In this article, we study the acoustoelectric effect as a result of the deformation potential mechanism of interaction~\cite{Chaplik1980}.
For this, we consider a monolayer TMDC, $\rm MoS_2$, exposed to a Rayleigh SAW. Rayleigh waves comprise of two (elastic) components of the force acting on the electrons corresponding to two components of the medium displacement vector. 
This is in contrast to the BG surface waves, which only have one component.
The displacement of the substrate medium due to the propagation of the Rayleigh SAW creates a strain field which results in the deformation potential and perturbs the electrons in $\mathrm{MoS_2}$.
From the general perspective, strain-induced perturbations serve an origin of a variety of interesting phenomena.
Therefore, they have been studied in systems like  graphene~\cite{Hsueaat9488,Farajollahpour2017}, TMDCs~\cite{Cazalilla}, Dirac~\cite{Araki2018} and Weyl semimetals~\cite{Cortijo2016}.

The standard acoustoelectric (AE) effect is associated with the transfer of SAW momentum to the electron subsystem resulting in a stationary electric current. Thus, the AE current is usually directed along the SAW wave vector reflecting the momentum transfer. 
The absence of inversion center in TMDCs results in the trigonal warping of the electron dispersion in the valleys. 
This property of Dirac materials leads to new transport phenomena, such as the photogalvainc effect when the system is exposed to external electromagnetic fields and to the emergence of AE current components perpendicular to SAW  direction. 

In this paper, we consider the effect of the trigonal warping of the electron dispersion in the TMDC and analyse its contribution to the Rayleigh SAW AE current.
We show, that depending on the direction of propagation of the SAW, we can distinguish between two dominant acoustoelectric currents: one conventional  current, which origin is diffusive, and the other Hall-like drift current due to the effect of the warping.
We analyse their properties in detail and compare them with the currents obtained in the case of piezoelectric interaction~\cite{Kalameitsev2019}.

The paper is organized as follows.
In Sec. II, we develop a formalism of the  deformation potential caused by the substrate displacement due to the Rayleigh SAW.
In Sec. III, we employ the Boltzmann transport theory to derive the expressions of the effective force acting on the electrons and electric currents.
In Sec. IV, we analyse the expressions of the currents and discuss their contributions.
Section V contains the analysis of results and the discussions.

\begin{figure}[tbp]
\includegraphics[width=0.49\textwidth]{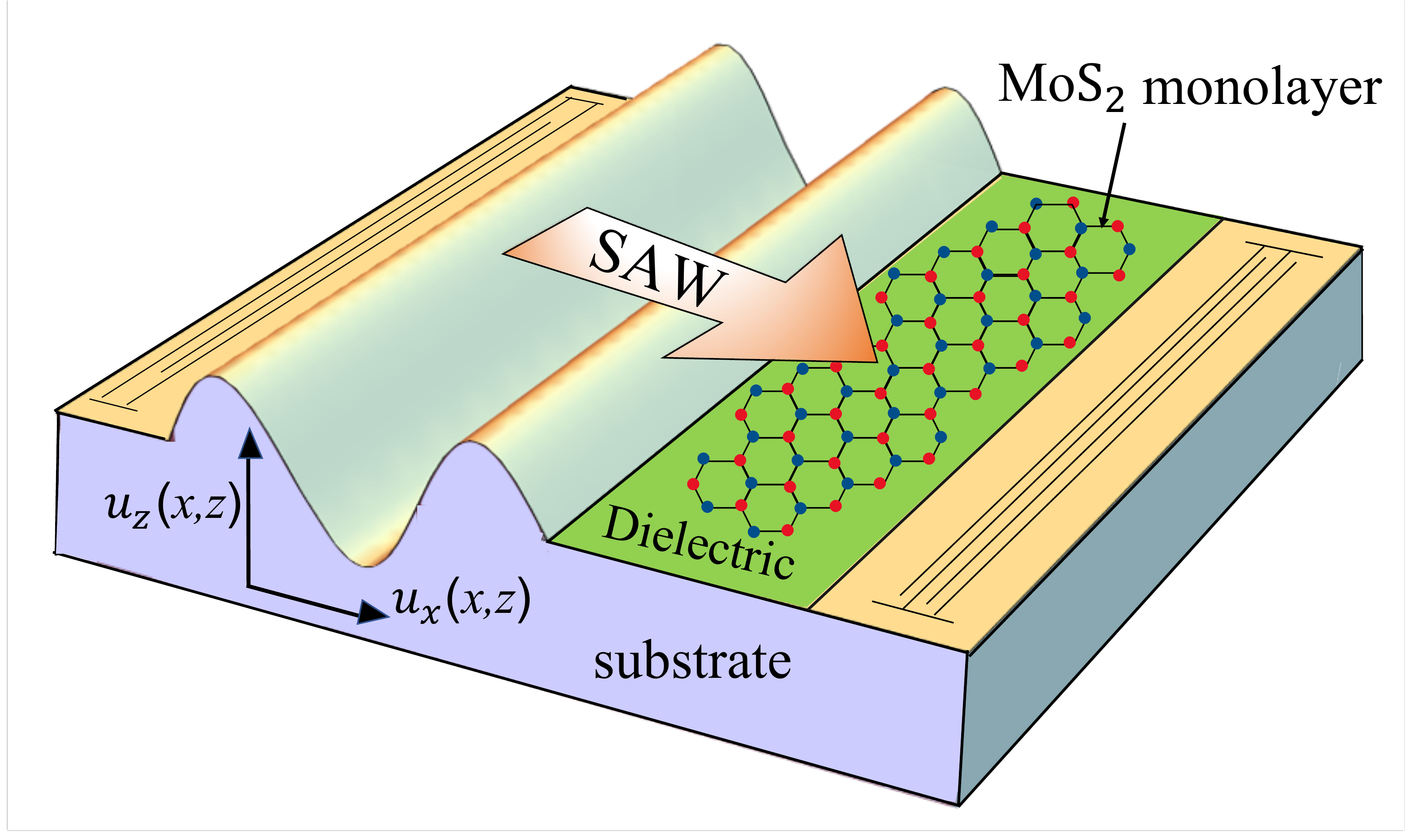}
\caption{Schematic illustration of the system: A monolayer of TMDC ($\rm MoS_2$) placed on an isotropic non-piezoelectric substrate  seperated by a dielectric layer. Propagation of Rayleigh SAW creates displacement of the substrate($u_x$ and $u_z$).}
\label{Fig1}
\end{figure}
%-----------------
%-----------------
%-----------------

%-----------------
%-----------------
%-----------------

\section{Deformation potential interaction of electrons with Rayleigh surface waves}
\subsection{Stress-tensor in the field of a Rayleigh wave}
We consider a system  which consists of a TMDC monolayer located on a semi-infinite substrate made of an isotropic material (Fig.~\ref{Fig1}).  
This assumption (of the isotropy of the substrate) simplifies the calculations in the meantime reflecting all the main properties of AE effect in the case of Rayleigh SAWs. 

The substrate displacement vector $\mathbf{u}$, which describes the propagation of the Rayleigh SAW along the surface of an isotropic medium, satisfies the equation~\cite{Boev2015,Landau}
\begin{equation}
\mathbf{\Ddot{u}} = c_t^2 \Delta \mathbf{u} + (c_l^2 - c_t^2) \rm grad~\rm div~\mathbf{u},
\end{equation}
where $c_l$ and $c_t$ are longitudinal and transverse sound velocities, respectively.
In the case of a Rayleigh SAW propagating along the x-direction of the $\textit{xy}$ plane, the components of the displacement vector read
\begin{equation}\label{displacement}
\begin{split}
u_z(x,z) &=  u_z(z) e^{ikx-i\omega t}, \\
u_x(x,z) &=  u_x(z) e^{ikx-i\omega t},\\
u_y &= 0,
\end{split}
\end{equation}
where
\begin{equation}
\label{displacement_2}
\begin{split}
u_z(z) &=  -i\kappa_lBe^{{\kappa}_lz} - ikAe^{{\kappa}_tz}, \\
u_x(z) &=  kBe^{\kappa_lz} + \kappa_tAe^{\kappa_t z},\\
\end{split}
\end{equation}
\begin{equation}
k_l = \sqrt{k^2 - \omega^2/c_l^2},~~ k_t = \sqrt{k^2 - \omega^2/c_t^2}.
\end{equation}
The exponential decay of the displacement vector components in $z$ direction reflects the surface nature of the Rayleigh wave, whereas the two other contributions in Eq.~\eqref{displacement_2} describe its two-component structure. 
The parameters A and B in Eq.~\eqref{displacement_2} are amplitudes, which can be related to each other using the boundary conditions on the surface of the substrate,
\begin{equation}
    B/A = -2\sqrt{1-\xi^2}/(2-\xi^2).
\end{equation}
Thus, they are dependent. The magnitudes of $A$ and $B$ are fixed by the source of Rayleigh waves. 
Hence, can express $A$ and $B$ through the SAW intensity,
\begin{equation}
    I_0 = c_t \xi \rho \int\limits_{-\infty}^{0}\Big(|\Dot{u}_x(x,z)|^2 + |\Dot{u}_z(x,z)|^2\Big)dz,
\end{equation}
yielding
\begin{equation}
A = \frac{\sqrt{I_0}  }{\omega \sqrt{c_t \xi\rho\kappa}}, B = -\frac{2\sqrt{1-\xi^2}}{(2-\xi^2)}\frac{\sqrt{I_0} }{\omega \sqrt{c_t \xi\rho\kappa}},
\end{equation}
where
\begin{equation}
\kappa = \frac{2(1-\xi^2)}{k_l(2-\xi^2)^2}(k_l^2 + k^2) + \frac{k^2 + k_t^2}{2k_t} - \frac{4\sqrt{1-\xi^2}k}{(2-\xi^2)}.
\end{equation}
Here $\xi$ is a constant characterizing the SAW dispersion such that $\omega = c_t \xi k$, and $I_0$ is the SAW intensity in W/m and $\rho$ is the density of the substrate material in kg/m$^2$.

The strain tensor is given by~\cite{Landau}
\begin{equation}\label{strain_tensor}
u_{\alpha\beta} =  \frac{1}{2}\bigg[\frac{\partial u_{\alpha}}{\partial x_{\beta}} + \frac{\partial u_{\beta}}{\partial x_{\alpha}} + \frac{\partial u_\gamma}{\partial x_\alpha}\frac{\partial u_\gamma}{\partial x_\beta}\bigg],
\end{equation}
where $x_i$ denote the coordinates, and $u_{\alpha}$ denote the displacement vector components. 
The non-zero components of the strain tensor calculated using~\eqref{strain_tensor} for the displacements given by Eq.~\eqref{displacement} are
\begin{equation}\label{strain}
\begin{split}
u_{zz} &= \frac{\partial u_z(x,z)}{\partial z} = u'_z(z) e^{ikx-i\omega t},
\\
u_{xx} &= \frac{\partial u_x(x,z)}{\partial x} = ik u_x(z) e^{ikx-i\omega t},
% \\u_{zx} &= u_{zx} = \frac{1}{2}\big( u_z(x,z)ik + (kBe^{k_lz}k_l \\
% & + k_t^2 A e^{k_tz})e^{ikx-i\omega t}\big)
\end{split}
\end{equation}
where $u_z'(z) = \partial(u_z)/\partial z$.
Some of the other components ($u_{yy}$, $u_{xy}$, $u_{yx}$) are zero.
Other (like $u_{xz}$) are finite but we do not use them.
It should be noted, that in the case of the SAW propagating in $y$ direction, $u_{xx}=0$, and then we should use $u_{yy}$ instead of $u_{xx}$ (the mathematical expression will be same, just $y$ replaced by $x$).

\subsection{Quasiclassical electron energy in the field of a Rayleigh wave}
Let us now derive the effective force acting on electrons due to the propagating Rayleigh wave. We start from the Hamiltonian of the system in the form,
\begin{equation}
\label{Hamiltonian_eff}
h_{\rm eff} = h_0 + h_{\rm strain},
\end{equation}
where $h_0$ is the bare Hamiltonian of the 2D TMDC and $h_{\rm strain}$ is the Hamiltonian reflecting the external perturbation due to the presence of the SAWs.
For a single-layer TMDC, we can write (in the continuos limit),
\begin{equation}\label{H_MOS2}
h_0 = \frac{\Delta}{2}\sigma^z
+v_0(\eta\sigma^x p_x + \sigma^yp_y),
%+ \frac{\lambda_{SO}\eta s^z}{2}(1-\sigma^z),
\end{equation}
where $p_x$, $p_y$ are the components of the electron momentum, $\eta = \pm 1$ is the valley index, $\sigma^{\alpha}$ are the Pauli matrices;
$\Delta$ is the band gap, and $v_0 = t_0a/\hbar$ with $a$ the lattice parameter, $t_0$ the hopping parameter. %$s^z$ is the spin projection along the z axis and $\lambda_{SO}$ accounts for spin-orbit coupling.

Following~\cite{Cazalilla}, we write (to the leading order, assuming $|u_{\alpha\beta}|\ll1$),
\begin{equation}\label{xx}
\begin{split}
h_{\rm strain}& = \beta_0 t_0 \sum_{\alpha}u_{\alpha\alpha} + \beta_1 t_0 \sum_{\alpha} u_{\alpha \alpha}\sigma^z\\
 &  + \beta_2 t_0[(u_{xx}-u_{yy})\sigma^x - 2u_{xy}\eta\sigma^y],
\end{split}
\end{equation}
where $\beta_0,\beta_1$ and $\beta_2$ are the Gr\"uneisen parameters \cite{Cazalilla}.
%given by $\beta_i = - \partial (\ln |t_i|)/\partial (\ln a)$ where {\it i} corresponds to atomic sites.% 

The terms of the Hamiltonian Eqs.~\eqref{H_MOS2} and~\eqref{xx} are given in the sublattices representaion since the TMDC lattice can be considered as two triangle sub-lattices inserted into each other and this representation is convenient. 
However, to study the electron transport, it is more practical to switch to the conduction and valence bands representation (cv-basis). The unitary transformation of bare Hamiltonian~\eqref{H_MOS2} into the cv-basis reads~\cite{Kovalev_2018},
\begin{equation}
\label{unitary}
\hat{U}  = \begin{pmatrix}
\cos{(\theta/2)}&\sin{(\theta/2)}\\
\sin{(\theta/2)}e^{i\eta\phi}&-\cos{(\theta/2)}e^{i\eta\phi}
\end{pmatrix},
\end{equation}
\begin{equation}
\nonumber
\hat{U}^+  = \begin{pmatrix}
\cos{(\theta/2)}&\sin{(\theta/2)}e^{-i\eta\phi}\\
\sin{(\theta/2)}&-\cos{(\theta/2)}e^{-i\eta\phi}\\
\end{pmatrix},
\end{equation}
where $\phi = \arctan(p_y/p_x)$ and $\theta$ is the polar angle,
\begin{equation}
\cos{\theta} = \frac{\Delta/2}{|\epsilon_{\eta s}|},~~\sin{\theta} = \frac{\eta v_0 p}{|\epsilon_{\eta s}|},
\end{equation}
where $\epsilon_{\eta s} = \sqrt{v_0^2 p^2 + \Delta^2/4}$.
Applying this unitary transformation to Eq.~\eqref{Hamiltonian_eff} we find
\begin{equation}
{H_0} = U^{+}h_0U = \begin{pmatrix}
\varepsilon_c^{\eta}&0\\
0&\varepsilon_v^{\eta}\\
\end{pmatrix},
\end{equation}
where $\epsilon_c^\eta$ and $\epsilon_v^\eta$ are the electron energies in conduction and valence bands given by
\begin{equation}\label{band_energy}
\varepsilon_{c,v}^{\eta} = \frac{1}{2}\left(s^z\eta\lambda_{SO} \pm \sqrt{4v^2p^2 + \Delta^2/4}\right),
\end{equation}
where ${p} = \sqrt{p_x^2 + p_y^2}$ is the absolute value of electron momentum. 

Furthermore, we assume that the TMDC layer is n-doped and the electrons in the conduction band form a degenerate electron gas with the Fermi energy $\mu$ and Fermi momentum $p_F$. Strictly speaking, the unitary transformation~\eqref{unitary} cannot be directly applied to the strain Hamiltonian since the latter depends on time and position in space, and thus it does not conserve the electron momentum and energy. 
Indeed, in general case, the electron scattering is inelastic, and the strain Hamiltonian should be included in the collision integral within the framework of Boltzmann  transport theory.  

However, the unitary transformation can be used in the \textit{quasiclassical} approximation approach. The typical SAW frequencies $\omega$ are much smaller than the characteristic electron energy, $\omega\ll\mu$, and the SAW wave vector is much smaller than the electron Fermi momentum, $k\ll p_F$. 
As such, the SAW wave can be treated as a \textit{weakly alternating in space and time classical field}, resulting in a classical potential force, acting on electrons. 
Then, this force can be written in the l.h.s. part of the Boltzmann equation describing the interaction with external fields.

Now, treating the strain term in the Hamiltonian as a quasi-static and quasi-uniform, and applying the unitary transformation~\eqref{unitary}, we find
\begin{equation}
{H}_{\rm strain} =     U^{+}h_{\rm strain}U = \begin{pmatrix}
\epsilon^{cc}&\epsilon^{cv}\\
\epsilon^{vc}&\epsilon^{vv}\\
\end{pmatrix},
\end{equation}
where $\epsilon^{cc(vv)}$ and $\epsilon^{cv(vc)}$ denote the intraband and interband electron-SAW interaction matrix elements, correspondingly. The frequency of the SAW is much smaller than the bandgap of $\rm MoS_2$, $\omega\ll \Delta$, thus it is possible to consider only the conduction band elements as a potential energy correction to the electrons in the conduction band. 
Performing the calculations, we find
%where $\beta_c = \beta_0 + \beta_1$ and $\beta_v = \beta_0 - \beta_1$\\
%
\begin{eqnarray}
\label{interaction_energy}
\epsilon^{cc} = t_0\bigg(\sum_{\alpha}u_{\alpha\alpha}\beta_0 + \sum_{\alpha}u_{\alpha\alpha}\beta_1\cos{\theta}\\
\nonumber
+ (u_{xx}-u_{yy})\beta_2 \cos{(\eta\phi)}\sin{\theta}\bigg).
\end{eqnarray}
This expression accounts for  two possible directions of the SAW:
If the wave propagates in x-direction, then $u_{yy}=0$; if, instead, the wave propagates in y-direction, $u_{xx}=0$.

It should also be noted, that formally, expression~\eqref{interaction_energy} contains both the electron momentum (via terms containing $\cos\theta$ and $\sin\theta$) and the electron position (via the components of the stress tensor). 
Nevertheless, since we work in the framework of the quasiclassical representation, and thus the position and momentum can be defined simultaneously,  Eq.~\eqref{interaction_energy} does not require the symmetrization procedure, as it is usually the case in the quantum description. Finally, the potential force acting on the conducting electrons due to the presence of the SAW reads a standard expression, $\textbf{F}(\textbf{r},t)=-\nabla\epsilon^{cc}(\textbf{r},t)$, and it does depend on the electron momentum $\textbf{p}$.

%

%-----------------
%-----------------
%-----------------

\section{The Boltzmann Transport theory for AE Current}
The Boltzmann equation reads
\begin{equation}\label{BE_transport}
    \frac{\partial f}{\partial t} + {\bf {\dot{p}}}\cdot\frac{\partial f}{\partial \bf{p}} + {\bf{\dot{r}}}\cdot\frac{\partial f}{\partial \bf{r}} = \frac{-(f - \langle f \rangle)}{\tau},
\end{equation}
where $f$ is the electron distribution function, which can be written as an expansion, $f = f_0 + f_1 + f_2 + O(3)$.
Here $f_0$ is the electron equilibrium distribution given by $f_0 = (\rm exp{[\varepsilon_{\bf p}-\mu(n)]/T} + 1)^{-1}$, where $\varepsilon_{\bf p}$ is given by the unperturbed energy of the conduction band in \eqref{band_energy}, and performing a series expansion in the limit $vp/\Delta \ll 1$ we write it in the following form, 
\begin{equation}\label{unperturbed energy}
    \varepsilon_{\bf p} = \frac{\Delta}{2} + \frac{ p^2}{2m^*}, \hspace{2cm} m^* = \frac{\Delta}{2v_0^2}.
\end{equation}
%
%where we have neglected the dependence of $\lambda_{SO}$.
%\vspace{0.6cm}
%
Furthermore, $\langle f \rangle$ is the locally equilibrium distribution function in the reference frame moving with the SAW. It depends on the local electron density $N(\bf r,t)$ via the chemical potential $\mu = \mu(N)$.
We expand $N(\bf r,t)$ in series, $N({\bf r},t) = n + n_1({\bf r},t) + n_2({\bf r}, t) + O(3)$, where $n$ is the unperturbed electron density and $n_i$ are the corrections to the density fluctuations. Also, $ \langle f \rangle ~ =~ f_0 + (n_1 + n_2 + ....)\partial_n f_0 + (n_1 + n_2 + ..)^2 \partial^2 f_0/\partial n^2 /2 $. 
Substituting these expansions in~\eqref{BE_transport}, the RHS of~\eqref{BE_transport} reads
\begin{equation}
\nonumber
-\frac{1}{\tau}\bigg(f_1 + f_2 - n_1\frac{\partial f_0}{\partial n} - n_2 \frac{\partial f_0}{\partial n} - \frac{n_1 n^*_1}{4}\frac{\partial^2 f_0}{\partial n^2} \bigg).
\end{equation}

The first-order corrections to the distribution function and the electron density read $f_1({\bf r},t) = (f_1 e^{i{\bf k.r}-i\omega t} + f^*_1 e^{-i{\bf k.r} + i\omega t})/2$ and $n_1({\bf r},t) = (n_1 e^{i{\bf k.r}-i\omega t} + n^*_1 e^{-i{\bf k.r} + i\omega t})/2$, respectively.

The second term in Eq.~\eqref{BE_transport} contains $\mathbf{\dot{p}}$, given by
\begin{eqnarray}
\nonumber
    \mathbf{\dot p} &=& \mathbf{\Tilde{F}(p)} + e\mathbf{\Tilde{E}}^i = \frac{1}{2}\Bigg({\mathbf {F(p)}} e^{i{\bf k.r}-i\omega t} + {\mathbf {F^*(p)}} e^{-i{\bf k \cdot r} + i\omega t}\Bigg)\\
    \label{p_dot}
   && + \frac{1}{2}\Bigg(e\mathbf{E}^ie^{i{\bf k.r}-i\omega t} + e\mathbf{E}^{i*} e^{-i{\bf k.r} + i\omega t}\Bigg),
\end{eqnarray}
%
% \textcolor{red}{where $\mathbf{{{F}}(p)} = -\nabla \epsilon_{cc}$. IS: Please, check!}
where $\mathbf{{\Tilde{F}}(p)} = -\nabla \epsilon^{cc}$ and 
${\bf \Tilde{E}}^i$ is the induced electric field due to the fluctuations of the electron density.
%Since $\mathbf{F(p)} = (F(\mathbf{p})\hat{x},0,0)$,

We find the components of $\mathbf{F}$ on the surface $z = 0$,
\begin{gather}\label{F_x}
    F_x({\bf p}) =  t(\beta_0 + \beta_1 \cos{\theta_p})(k^2 u_x(0)-iku_z'(0)) \\
    \nonumber
    + t\beta_2 \cos({\eta \phi}) \sin{\theta_p} k^2 u_x(0),
\end{gather}
\begin{gather}\label{F_y}
    F_y({\bf p}) =  t(\beta_0 + \beta_1 \cos{\theta_p})(k^2 u_y(0) - iku_z'(0))\\
    \nonumber
    - t\beta_2 \cos({\eta \phi}) \sin{\theta_p} k^2 u_y(0),
\end{gather}
and the complex conjugated parts read
\begin{gather}\label{F_x_conjugate}
    F_x^*{({\mathbf p})} = t(\beta_0 + \beta_1 \cos{\theta_p})(k^2 u^*_x(0)+ iku^{*'}_z(0))\\
    \nonumber
    + t\beta_2 \cos({\eta \phi}) \sin{\theta_p} k^2 u^*_x(0),
\end{gather}
\begin{gather}\label{F_y_conjugate}
    F_y^*{({\mathbf p})} = t(\beta_0 + \beta_1 \cos{\theta_p})(k^2 u^*_y(0)+ iku^{*'}_z(0))\\
    \nonumber
    - t\beta_2 \cos({\eta \phi}) \sin{\theta_p} k^2 u^*_y(0).  
\end{gather}
Combining the l.h.s. and r.h.s terms in the first order gives
\begin{gather}\label{first_order}
    -i(\omega - {\bf k}\cdot\mathbf{v})f_1 + ({\bf{F(\bf p)}} + e{\bf E}^i)\cdot\frac{\partial f_0}{\partial {\bf p}}=
    \\
    \nonumber
    =-\frac{1}{\tau}(f_1 - n_1\frac{\partial f_0}{\partial n}),
\end{gather}

We find
\begin{eqnarray}\label{f_1}
    f_1 = \frac{n_1 ({\partial f_0}/{\partial n}) - ({ F(\bf p)} + e{\bf E}^i)(\partial f_0/\partial \bf p)\tau}{1-i(\omega - \bf k \cdot v)\tau}.
\end{eqnarray}

Furthermore, we employ the continuity equation, $\partial \rho/ \partial t = - \nabla\cdot\bold{j}$, which gives
\begin{equation}\label{continuity_eq}
    e \omega n_1 = {\bf k}\cdot\mathbf{j}.
\end{equation}
Then, using
\begin{eqnarray}
    n_1 = \int \frac{d{\bf p}}{(2\pi)^2} f_1,                       ~~~\mathbf{j} = e \int \frac{d{\bf p}}{(2\pi)^2}{\bf v}f_1,
\end{eqnarray}
and substituting~\eqref{f_1}, Eq.~\eqref{continuity_eq} becomes
\begin{gather}
    e \omega n_1 = e {\bf k}\int \frac{d{\bf p}}{(2\pi)^2} {\bf v} \frac{n_1 ({\partial f_0}/{\partial n})}{1-i(\omega - \mathbf {k}\cdot\mathbf{v})\tau}-\\
    \nonumber
    -\frac{(\mathbf{F(p)} + e\mathbf{E}^i)(\partial f_0/\partial \bf p)\tau}{1-i(\omega - \mathbf {k}\cdot\mathbf{v})\tau}.
\end{gather}
We should note here that
\begin{eqnarray}\label{derivatives}
    {\bf v}\frac{\partial f_0}{\partial n} = {\bf v}\frac{\partial \mu}{\partial n}\frac{\partial f_0}{\partial n} = -\frac{\partial \mu }{\partial n}{\bf v}\frac{\partial f_0}{\partial \varepsilon_{\bf p}} = - \frac{\partial \mu}{\partial n}\frac{\partial f_0}{\partial {\bf p}}.
\end{eqnarray}
Taking into account these equalities and introducing the diffusion vector
\begin{eqnarray}
    {\bf R} = \int \frac{d{\bf p}}{(2\pi)^2} \frac{{\bf v}\cdot(-\partial f_0/\partial \epsilon_{\bf p})}{1-i(\omega - \bf k\cdot v)\tau} \frac{\partial \mu}{\partial n},
\end{eqnarray}
and the conductivity tensor
\begin{eqnarray}\label{conductivity}
    \sigma_{\alpha \beta} = e^2 \tau \int \frac{d {\bf p}}{(2\pi)^2} \frac{v_{\alpha}v_{\beta}}{1-i(\omega - {\bf k \cdot v})\tau} \bigg(-\frac{\partial f_0}{\partial \varepsilon_{\bf p}}\bigg),
\end{eqnarray}
we find
\begin{eqnarray}\label{n_1}
    n_1 = \frac{k_{\alpha}\sigma_{\alpha\beta}(E_{\beta}^i + F_{\beta}(p)/e)}{e(\omega - {\bf k\cdot R})}.
\end{eqnarray}

The induced electric field obeys the Maxwell's equation the solution of which reads
\begin{equation}\label{Maxwell}
    {\bf E}^i = -4 \pi i e {\bf k} n_1/(k\varepsilon + k).
\end{equation}
Substituting~\eqref{Maxwell} in Eq.~\eqref{n_1} yields
\begin{equation}\label{n1}
    n_1 = \frac{k_{\alpha} \sigma_{\alpha \beta} F_{\beta}(p)}{e (\omega - {\bf k \cdot R}) g({\bf k},\omega)},
\end{equation}
where
\begin{equation}\label{g}
    g({\bf k},\omega) = 1 + i\frac{4\pi}{\epsilon + 1}\frac{k_{\alpha}\sigma_{\alpha\beta}k_{\beta}}{k(\omega - {\bf k \cdot R})}
\end{equation}
is the dielectric function of the 2DEG.

Collecting the second-order terms in both the l.h.s. and r.h.s. of \eqref{BE_transport} gives
\begin{gather}\label{second_order}
        ({{\bf F}^*(\bold{p})} + e {{\bf E}^{i*}})\frac{\partial f_1}{\partial {\bf p}}  + ({{\bf F(p)}} + e{\bf E}^i)\frac{\partial f_1^*}{\partial {\bf p} }=\\
        \nonumber= -\frac{1}{\tau}(f_2 -
        \overline{n}_2\frac{\partial f_0}{\partial n} - \frac{n_1 n_1^*}{4}\frac{\partial^2 f_0}{\partial n^2}),
\end{gather}
where we have omitted the fast-oscillating terms containing $e^{2i\omega t}$ since they vanish after the time averaging.

%-------------------
%-------------------
%-------------------
\begin{figure}[!b]
\includegraphics[width=0.47\textwidth]{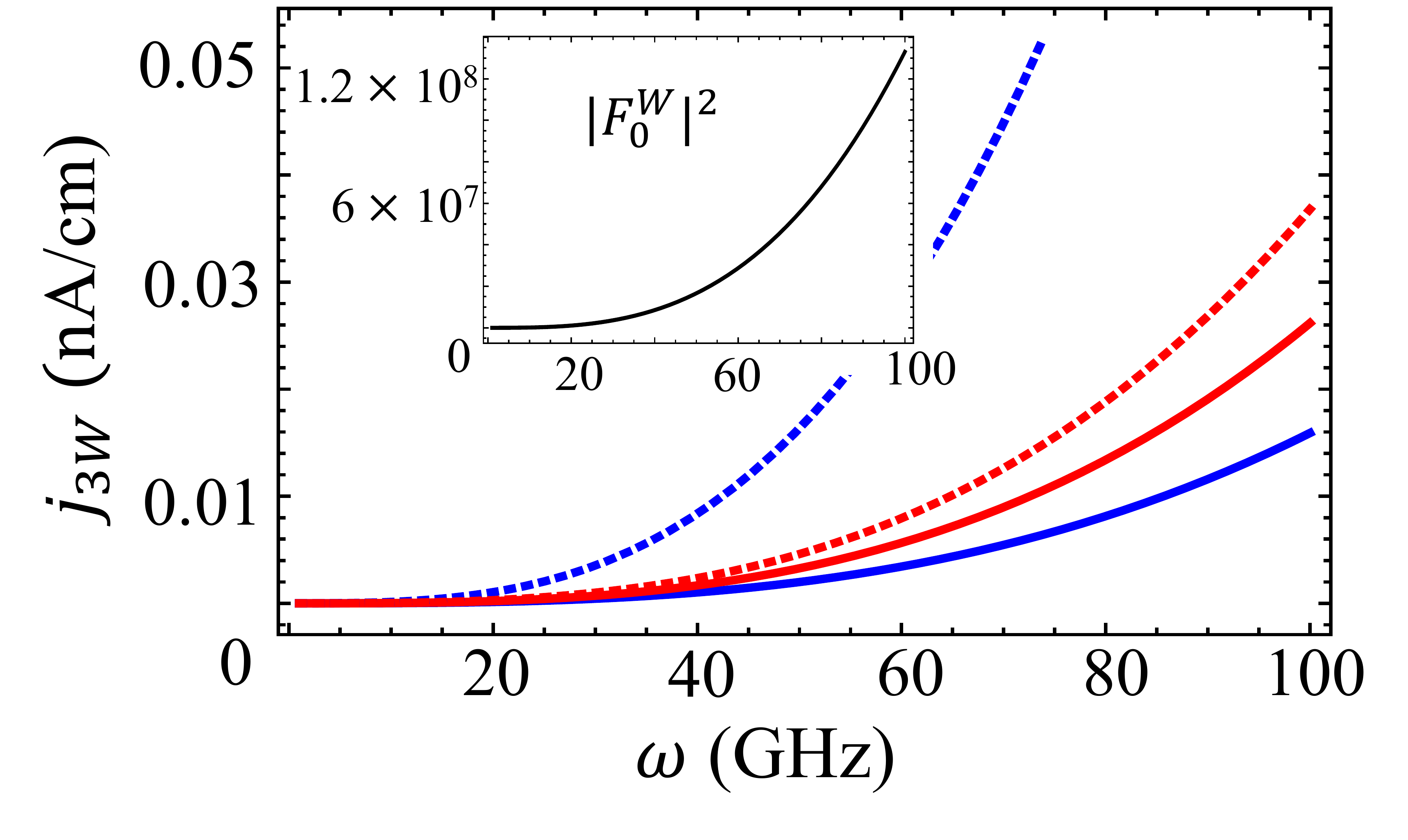}
\caption{Spectrum of  warping current(b) for different values of $n$ for two fixed values of $\tau$: $10^{-14}$s~    (solid) and $5 \times10^{-14}$ s~(dotted). The blue and red curves represent $n = 1 \times 10^{10} \rm cm^{-2}$ and $ n = 5 \times10^{10} \rm cm^{-2}$ .}
    \label{Fig2}
\end{figure}
% %

\begin{figure}[tbp]
\includegraphics[width=0.49\textwidth]{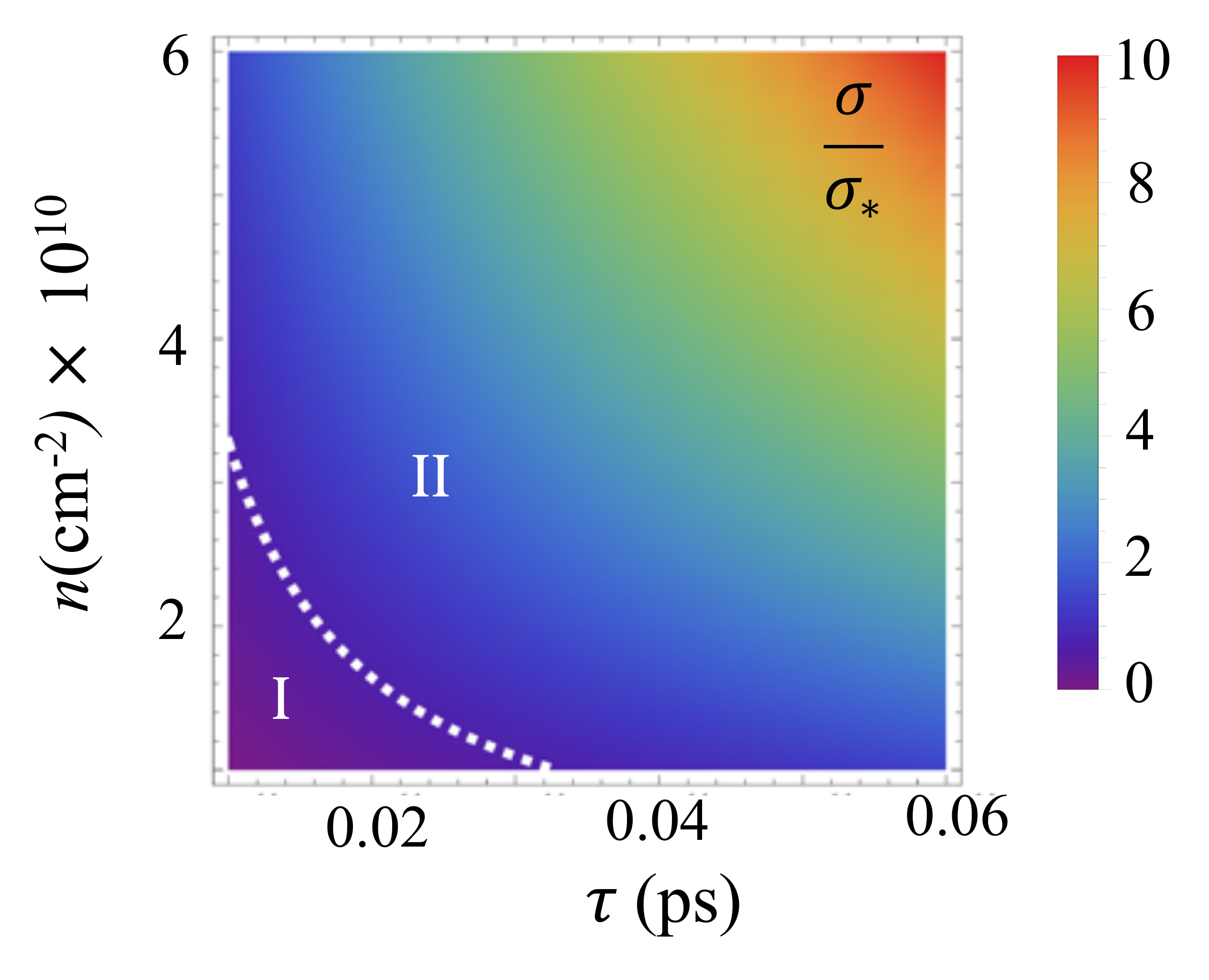}
\caption{Density plot for $\sigma/\sigma_*$. Dashed white contour depicts $\sigma/\sigma_*=1$ which divides the acoustoelectric effect in two regimes: I. low-$n\tau$ regime where $\sigma/\sigma_* \ll 1$ and II. high-$n\tau$ regime where $\sigma/\sigma_* \gg 1$.}
    \label{Fig3}
\end{figure}
\begin{figure*}%[tbp]
\includegraphics[width=0.9\textwidth]{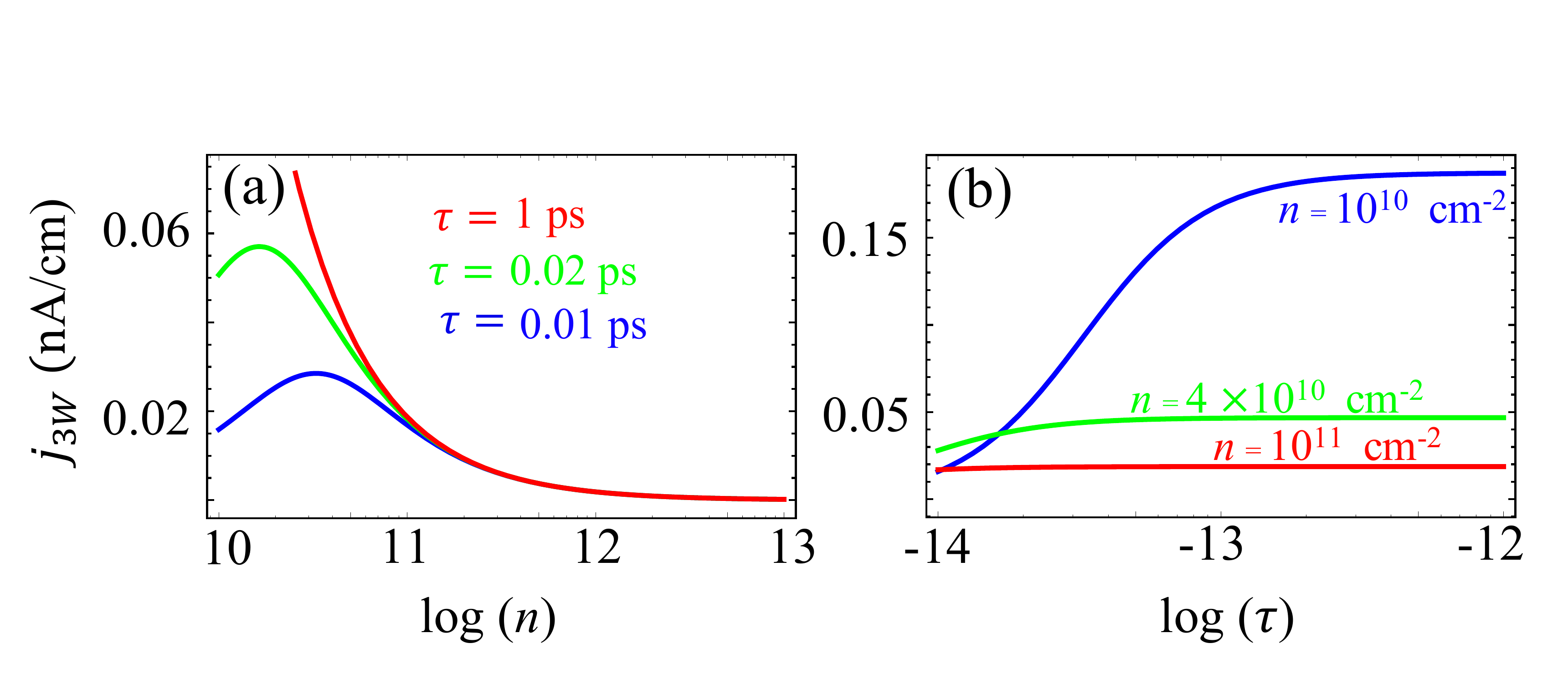}
\caption{Warping current (a) as a function of $n$ with fixed $\tau$ (b) as a function of $\tau$ with $n$ fixed. The valley index has been taken as $\eta = +1$ and $\omega =  10^{11}$ Hz.}
    \label{Fig4}
\end{figure*}

%
%
%

% %
% %
% %

%-------------
%-------------
%-------------

\section{Stationary electric current}

The finite stationary electric current can be found in the second-order with respect to external drag force,
\begin{equation}\label{2_order_current}
    {\bf j} = e \int {\bf dp} f_2 {\bf v}/(2\pi)^2.
\end{equation}
Noting that the second and third terms in the r.h.s of Eq.~\eqref{second_order} do not contribute to this current, we write
\begin{eqnarray}\label{f_2}
    f_2  =  -\tau\bigg[({{\bf F}^*(\bold{p})} + e {{\bf E}^{i*}})\frac{\partial f_1}{\partial {\bf p}}  + ({{\bf F(p)}} + e{\bf E}^i)\frac{\partial f_1^*}{\partial {\bf p} }\bigg],~~~~
\end{eqnarray}
thus disregarding the other terms in $f_2$ which are not interesting for us.
Substituting Eq.~\eqref{f_2} in~\eqref{2_order_current}, we find
\begin{eqnarray}
    j_{\alpha} = -\frac{e^2 \tau}{2} {\rm Re} \int \frac{d{\bf p}}{(2\pi)^2}v_{\alpha}\left(\frac{F^{*}_{\beta}(\bold{p})}{e} + E^{i*}_{\beta}\right)\frac{\partial f_1}{\partial p_{\beta}}.
\end{eqnarray}
Integrating by parts yields
\begin{gather}
\label{current_1}
    j_{\alpha} = \frac{e^2 \tau}{2}\Big[{\rm Re} \int \frac{\bf dp}{(2\pi)^2} f_1 \frac{\partial v_{\alpha}}{\partial p_{\beta}}\left(\frac{{F}^*_{\beta}(\bold{p})}{e} + { E}_{\beta}^{i*}\right)+\\
    \nonumber
    + {\rm Re} \int \frac{\bf dp}{(2\pi)^2} f_1 v_{\alpha} \frac{\partial {F}^*_{\beta}(\bold{p})/e}{\partial p_{\beta}} \Big].
\end{gather}
Substituting $f_1$ from \eqref{f_1}, we can distinguish between several contributions to the electric current density,
\begin{eqnarray}\label{j_alpha}
    j_{\alpha} = j_{\alpha}^{(1a)} + j_{\alpha}^{(1b)} + j_{\alpha}^{(2a)} + j_{\alpha}^{(2b)},
\end{eqnarray}
the two of which are diffusive currents,
\begin{gather}
\label{j_1a}
 j_{\alpha}^{(1a)} = \frac{e\tau}{2}{\rm Re} \int \frac{d^2 p}{(2\pi)^2}\frac{k_{\gamma}\sigma_{\gamma\delta}(E_{\delta}^i + F_{\delta}(\mathbf p)/e)}{(\omega - {\mathbf{k \cdot R}})} \frac{\partial v_{\alpha}}{\partial p_{\beta}}\times\\
 \nonumber
 \times\left(E_{\beta}^{i*} + \frac{F^*_{\beta}(\mathbf p)}{e}\right)\frac{(-\partial f_0/\partial \varepsilon_{\bf p})}{1-i(\omega - {\bf k \cdot v})\tau}\frac{\partial \mu}{\partial n},
\end{gather}
\begin{gather}
\label{j_2a}
    j_{\alpha}^{(2a)} = \frac{e\tau}{2}{\rm Re} \int \frac{ d^2p}{(2\pi)^2}
    \frac{k_{\gamma}\sigma_{\gamma\delta}(E_{\delta}^i + F_{\delta}(\bold{p})/e)}{(\omega - {\mathbf{k.R}})}\times\\
    \nonumber
    \times v_{\alpha}\frac{\partial F_{\beta}^*(\bold{p})}{\partial p_{\beta}}\frac{(-\partial f_0/\partial \varepsilon_{\bf p})}{1-i(\omega - {\bf k \cdot v})\tau}\frac{\partial \mu}{\partial n}
\end{gather}        
and the other two are drift currents,
\begin{gather}
\label{j_1b}
    j_{\alpha}^{(1b)} = \frac{e^3\tau^2}{2}{\rm Re} \int \frac{d^2p}{(2\pi)^2} v_{\gamma}\left(\frac{F_{\gamma}(\mathbf p)}{e}+E^i_{\gamma}\right)\frac{\partial v_{\alpha}}{\partial p_{\beta}}\\
    \nonumber
    \times\left(\frac{F_{\beta}^*(\bold{p})}{e} 
    + E^{i*}_{\beta}\right)\frac{(-{\partial f_0}/{\partial \varepsilon_{
    \bf p}})}{1-i(\omega - {\mathbf{k \cdot v}})\tau},
\end{gather}
\begin{gather}
\label{j_2b}
    j_{\alpha}^{(2b)} = \frac{e^3\tau^2}{2}{\rm Re} \int \frac{ d^2p}{(2\pi)^2}v_{\gamma}\left(\frac{F_{\gamma}(\bold{p})}{e} +
    E_{\gamma}^i\right)\times\\
    \nonumber
    \times v_{\alpha}\frac{\partial F_{\beta}^*(\bold{p})}{\partial p_{\beta}}\frac{(-\partial f_0/\partial \varepsilon_{\bf p})}{1-i(\omega - {\bf k \cdot v})\tau}.
\end{gather}

In what follows, we  consider the relevant experimental situation when $\omega\tau\ll1$ and $\mathbf{k}\cdot\mathbf{v}\tau\ll1$ in which the diffusive vector is small $R \simeq 0$.
Also, we will consider small temperatures, at which we can replace,
\begin{equation}\label{substitutions}
    \frac{\partial \mu}{\partial n} = \frac{\pi\hbar^2}{m^*}, \hspace{0.4cm}-\frac{\partial f_0}{\partial \varepsilon_p} = \delta(\mu-\varepsilon).
\end{equation}
%

%--------------------------------
%--------------------------------
%--------------------------------

%
%
%

\subsection{Drag electric current without the trigonal warping}
If we consider the electric current in the x direction when the electric field is also oriented in x direction,
the calculations show~\cite{SMBG}, that the biggest contribution in this case is given by the diffusive current term $j^{(1a)}$,
\begin{equation}\label{diffusive_j}
    j_{x}^{(1a)} = j_D = \frac{e\tau}{8m^*} \frac{k\sigma}{\omega}\frac{|F^d_0(\omega,n)|^2}{1+(\sigma/\sigma_*)^2}, 
\end{equation}
where
\begin{eqnarray}\label{amplitude_d}
    |F^d_0(\omega,n)|^2 = \frac{t^2}{e^2} \Bigg[|u_x(0)|^2k^4\Bigg(2\Big(\beta_0 + \frac{\beta_1}{\sqrt{\gamma^2 + 1}}\Big)^2 ~ \\
    \nonumber
    +\beta_2^2\frac{\gamma^2}{\gamma^2 + 1}\Bigg) + ~|u_z'(0)|^2k^2\Big(\beta_0 + \frac{\beta_1}{\sqrt{\gamma^2 + 1}}\Big)^2\Bigg].
\end{eqnarray}
Here  $\sigma={n e^2 \tau}/{m^*}$ is the Drude conductivity, $\gamma=\sqrt{\Delta(\mu(n) - \Delta/2)}/(\Delta/2)$, $\sigma_* = \varepsilon_0(\varepsilon+1)s/4\pi$, and $s = c_t\xi$.
If we consider the electric current in $y$ direction when the electric field is also in $y$ direction, we find similar results.

%-----------------------
%-----------------------
%-----------------------

\subsection{Drag electric current due to the trigonal warping contribution}

If we include the trigonal warping term~\cite{Chen:2020aa,Twarp2013} in the system Hamiltonian~\eqref{Hamiltonian_eff}, we find
\begin{equation}
    h_{\rm eff}= h_0 + h_{\rm strain} + h_{3W},
\end{equation}
where
\begin{equation}
h_{3W} =\frac{C}{\hbar^2}\begin{pmatrix}
0&p_{+}^2\\
p_-^2&0\\
\end{pmatrix}\hspace{0.5cm} {\rm with} \hspace{0.5cm} p_{\pm} = \eta p_x \pm i p_y,
\end{equation}
and $C$ is the warping strength in eV $\AA^2$.
Performing the derivations (see~\cite{SMBG}), we find the dispersion,
\begin{equation}\label{dispersion}
    \varepsilon_{\bf p} = \frac{\Delta}{2} + \frac{ p^2}{2m^*} + C' (p_x^3 - 3p_x p_y^2),
\end{equation}
where $C' = 2C\eta v_0/\hbar^2(\Delta/2)$.
Furthermore, we can estimate the electric current density due to the warping terms. 
They will enter Eqs.~\eqref{j_1a}-\eqref{j_2b} through the terms,
\begin{eqnarray}
\label{Eqvwarp}
v_x&=& \frac{p_x}{m} + 3C'(p_x^2-p_y^2),\\
\nonumber
v_y&=& \frac{p_y}{m} + 6C'p_xp_y,\\
\nonumber
\frac{\partial v_\alpha}{\partial p_\beta} &=& \begin{pmatrix}
\frac{1}{m^*} + 6C'p_x &\hspace{0.8cm} -6 C'p_y\\
\\
-6 C' p_y&\hspace{0.8cm} \frac{1}{m^*}-6 C' p_x\\
\end{pmatrix}.
\end{eqnarray}
As expected, they give a small correction to the main current.

%-----------------------------
%-----------------------------
%-----------------------------

\subsection{Hall-like currents}

%In this section, we calculate the hall current $j_x$ for field in $y$ direction.\\

If we take the force in $y$ direction, the current~\eqref{j_1b} in $x$ direction ($\alpha=x$) with account of~\eqref{Eqvwarp} yields~\cite{SMBG}
\begin{equation}\label{warping_j}
    j_{3W} = j_{x}^{(1b)} = \frac{-2e^3\tau^2\eta}{\pi}\frac{Cv_0m^*}{\hbar^2\Delta}\left(\mu(n)-\frac{\Delta}{2}\right)\frac{|F^W_0(\omega,n)|^2}{1+(\sigma/\sigma_*)^2},
\end{equation}
where
\begin{eqnarray}\label{amplitude_W}
    |F^W_0(\omega,n)|^2 = \frac{t^2}{e^2} \Bigg[|u_y(0)|^2k^4\Bigg(\Big(\beta_0 + \frac{\beta_1}{\sqrt{\gamma^2 + 1}}\Big)^2 ~ \\
    \nonumber
    +\frac{\beta_2^2}{8}\frac{\gamma^2}{\gamma^2 + 1}\Bigg) + ~|u_z'(0)|^2k^2\Big(\beta_0 + \frac{\beta_1}{\sqrt{\gamma^2 + 1}}\Big)^2\Bigg].
\end{eqnarray}
The other contributions from Eqs.~\eqref{j_1a},~\eqref{j_2a} and~\eqref{j_2b} vanish after the momentum and angle integrations.

%

%
%
%

%------------------
%------------------
%------------------

\section{Results and discussion}

Let us analyze the warping current~\eqref{warping_j} for the parameters characteristic of $\rm MoS_2$ deposited on an isotropic non-piezoelectric substrate such as silicon. 
In equilibrium, the net warping curent is zero since both the valleys give a contribution to the current equal in magnitude and different in sign. 
Thus, to observe a nonzero current, the time reversal symmetry breaking is required.
In experiments, it is achieved by illuminating the sample by a circularly-polarized light which selective pumps only one of the valleys and thus creates the electron population imbalance.

Let us discuss the parameters.
The typical electron densities range between $10^{10}-10^{13}$ $\rm cm^{-2}$, and the electron relaxation times are of the order of $10^{-13}-10^{-14}$~s~\cite{Saito2016}.
We will also use the warping constant $C=-1.02 $~eV $\AA^2$ calculated for MoS$_2$~\cite{Twarp2013}.
We note, that the Hall-like drift current~\eqref{warping_j} depends on the valley index (while the diffusive drag current is independent of it, see~\cite{[{See Supplemental Material at [URL]. It gives the details of the derivations of main formulas}]SMBG}).

First of all, the analysis of the forces~\eqref{amplitude_d} and~\eqref{amplitude_W} shows that they are only dependent on the frequency $\omega$ and are nearly independent of $n$ since $\gamma(n)$ is always smaller than unity for the chosen electron densities.
Therefore, the spectrum of the warping current presented in Fig.~\ref{Fig2} behaves similarly to the force squared $|F_0^W|^2$ (shown in the inset). 

We find that the warping current can reach several nA/cm. 
The diffusive drag current follows the same dependence on frequency of the Rayleigh SAW as the warping diffusive current but it is of higher magnitude (up to $\mu$A/cm). 
The difference in the magnitude of the two currents, as it follows from Eqs.~\eqref{diffusive_j} and~\eqref{warping_j}, is due to the presence of the warping constant in the warping current, and the ratio between the two currents remains constant.
Different values of $n$ and $\tau$ resuls in a shift of the curves in Fig.~\ref{Fig2} (keeping the behaviour of the spectrum unchanged). 

Furthermore, depending on the parameters $n$ and $\tau$, the ratio $\sigma/\sigma_*\sim n\tau$ can be smaller, comparable, or much greater than unity (see Fig.~\ref{Fig3}).
For realistic parameters, $\sigma/\sigma_*$ can take values up to  $\sim10^4$. However, it also can be small in disordered or low-doped samples.
It allows us to consider analytically the response of the system in two regimes of small and large $n\tau$. 
 %\textcolor{blue}{Mention conditions when electron density and lifetime can be low to these orders that we have taken? }

The corresponding expressions for diffusive and warping currents reads
\begin{equation}\label{j_D_approx}
j_D  = \left\{
        \begin{array}{ll}
            {(e^3 \tau^2 n/{s m^*})(F_0^{d})^2}, & \quad  \sigma/\sigma_* \ll 1\\
            \\
            {(\sigma_*^2/{ens})(F_0^{d})^2},  & \quad \sigma/\sigma_* \gg 1
        \end{array}
    \right.
\end{equation}

\begin{equation}\label{j_W_approx}
j_{3W}  = \left\{
        \begin{array}{ll}
            (-2Cv_0m^*/\pi \hbar^2\Delta)e^3\tau^2\\ ~~\times\left(\mu(n)-\frac{\Delta}{2}\right){(F^W_0)^2}, & \quad  \sigma/\sigma_* \ll 1\\
            \\
            (-2Cv_0m^*/\pi \hbar^2\Delta)\sigma_*^2/{en^2}\\
            ~~\times\left(\mu(n)-\frac{\Delta}{2}\right){( F_0^{W})^2}.  & \quad \sigma/\sigma_* \gg 1
        \end{array}
    \right.
\end{equation}
We see that the ratio of the diffusive (normal) and drift (warping) currents remains constant in both the limiting cases, as the functional form of the dependence on parameters $n$ and $\tau$ is the same. 

Figure~\ref{Fig4} shows the results of the numerical calculation of the electric current densities in general case.
%, we analyse this behaviour in different regimes by plotting the warping current as a logarithmic function of $n$ and $\tau$.
We see that when both $n$ and $\tau$ are  small, the current grows linearly with $n$ [panel (a)] and quadratically with $\tau$ [panel (b)] until it reaches a peak at intermediate values of $n\tau$.
With the further increase of electron density, the current starts to drop as $\propto n^{-2}$ at large $n$ [panel (a)] which is consistent with~\eqref{j_W_approx}.
With the increase of $\tau$ (keeping $n$ constant), the system switches to the high $n\tau$ regime and the current becomes independent of $\tau$.

It is also important to compare our results with the ones found for BG acoustic waves~\cite{Kalameitsev2019} when the 2D material is located on a piezoelectric substrate. 
We see that the currents due to the Rayleigh waves are smaller in magnitude for the same frequencies of SAW. 
The physics of acoustoelectric interaction remains similar in for both the waves. In particular, the diffusive current turns out to be the dominant in both the cases and the trigonal warping also provides qualitatively similar behavior. 
Thus, the general conclusion is that the drag field due to the deformation potential is weaker than the electric field provided by the piezoelectric surface.
However, in real samples both the acoustic waves coexist and give a combined effect on the electronic system. In samples without the piezoelectric substrate, the Rayleigh waves give the largest impact to acoustoelectric effect.

\section{Conclusions}

We have analysed the acoustoelectric effect due to the propagation of Rayleigh surface acoustic waves in monolayers of transition-metal dichalcogenides. Our calculations show that resulting current  comprise of two main contributions: the conventional diffusive current in the direction of the drag field and the Hall-like warping current in the direction perpendicular to the drag field. 
It emerges due to the trigonal warping of the electron dispersion and thus the orientation of the crystal with respect to the direction of propagation of the acoustic wave matters. We have analysed the dependence of the electric current density on the material parameters like the electron density and the relaxation time, and the SAW parameters such as its dispersion and intensity. 
We have also considered the regimes of small and large electron densities and electron scattering times. 
Our theory carries a potential for its application in experiments aimed at studying two-dimensiomnal Dirac materials using acoustic waves.

%\acknowledgements
%\section{acknowledgement}
KS and IGS acknowledge the support by the Institute for Basic Science in Korea (Project No.~IBS-R024-D1). AVK and VMK were supported by the Russian Foundation for Basic Research and Government of the Novosibirsk Region (Project No.~19-42-540011).

 \bibliography{reference}
\bibliographystyle{apsrev4-1}

\end{document}